\documentclass[amssymb,prd,superscriptaddress,aps,nofootinbib,twocolumn,showpacs]{revtex4-1}
\usepackage{graphicx, epsfig, amssymb} 
\usepackage{amsmath, amsfonts}
\usepackage{bm} 
\usepackage[breaklinks]{hyperref}
\usepackage{color}
\usepackage{enumerate}

\def\nn{\nonumber}
\DeclareMathOperator\erf{erf}

\def\be{\begin{equation}}
\def\ee{\end{equation}}
\def\beq{\begin{eqnarray}}
\def\eeq{\end{eqnarray}}
\def\DM{{\rm \mbox{{\tiny DM}}}}
\def\GC{{\rm \mbox{{\tiny GC}}}}
\def\DF{{\rm \mbox{{\tiny DF}}}}
\def\GW{{\rm \mbox{{\tiny GW}}}}

\def\MAX{{\rm \mbox{{\scriptsize max}}}}
\def\obs{{\rm \mbox{{\scriptsize obs}}}}

\def\INT{{\rm \mbox{{\scriptsize int}}}}

\def\apj{Astrophysical Journal }                
\def\apjl{Astrophysical Journal }             

\def\mnras{MNRAS }            
\def\nat{Nature }

\begin{document}

\title{Binary pulsars as dark-matter probes}

\author{Paolo Pani}\email{paolo.pani@roma1.infn.it}
\affiliation{Dipartimento di Fisica, ``Sapienza'' Universit\`a di Roma \& Sezione INFN Roma1, Piazzale Aldo Moro 5, 00185, Roma, Italy.}
\affiliation{CENTRA, Departamento de F\'{\i}sica, Instituto Superior T\'ecnico, Universidade de Lisboa, Avenida~Rovisco Pais 1, 1049 Lisboa, Portugal.}

\begin{abstract} 
During the motion of a binary pulsar around the Galactic center, the pulsar and its companion experience a wind of dark-matter particles that can affect the orbital motion through dynamical friction. We show that this effect produces a characteristic seasonal modulation of the orbit and causes a secular change of the orbital period whose magnitude can be well within the astonishing precision of various binary-pulsar observations. 
Our analysis is valid for binary systems with orbital period longer than a day. 
By comparing this effect with pulsar-timing measurements, it is possible to derive model-independent upper bounds
on the dark-matter density at different distances $D$ from the Galactic center. For example, the precision timing of J1713+0747 imposes $\rho_{\DM}\lesssim 10^5\,{\rm GeV/cm}^3$ at $D\approx7\,{\rm kpc}$. The detection of a binary pulsar at $D\lesssim 10\,{\rm pc}$ could provide stringent constraints on dark-matter halo profiles and on growth models of the central black hole. The Square Kilometer Array can improve current bounds by $2$ orders of magnitude, potentially constraining the local density of dark matter to unprecedented levels.
\end{abstract}


\maketitle

\section{Introduction}\label{sec:intro}

Dark matter (DM) is five times as abundant as baryonic matter in the Universe but its properties remain mysterious. While there is strong evidence for DM particles to be nonrelativistic, their mass, spin, charges, and annihilation and interaction cross sections are unknown~\cite{BertoneBook}. The plethora of DM candidates makes direct detection extremely challenging and model dependent.
A further astrophysical problem related to DM is to measure its density profile in the Milky Way. Conventional cold DM cosmological models predict cuspy halos which might give rise to strong gamma-ray emission through DM annihilation. However, such DM cusps seem to be in conflict with observations (see e.g. Refs.~\cite{1994ApJ...427L...1F,1994Natur.370..629M}). 

In this intricate scenario, the universality of the gravitational interactions might represent a stronghold for model-independent tests of DM. In this paper we point out that binary pulsars are the ideal laboratory for such gravitational tests. Pulsar-timing techniques allow us to measure some of the orbital parameters with exquisite precision, and are routinely used to measure the mass of neutron stars (e.g. Refs.~\cite{Demorest:2010bx,Antoniadis26042013}) and to perform some of the most stringent tests on Einstein's theory of general relativity, as in the case of the celebrated Hulse-Taylor binary pulsar~\cite{Hulse:1974eb} and of the more recent double-pulsar system~\cite{Lyne:2004cj} (see also Ref.~\cite{2012MNRAS.423.3328F} and Refs.~\cite{Kramer:2004hd,Yunes:2013dva,Berti:2015itd} for some reviews).

During the motion of a binary pulsar around the Galactic center, the pulsar and its companion experience a wind of DM particles similar to the one that produces an annual modulation in the scattering rate in direct-detection experiments due to Earth's motion around the Sun~\cite{PhysRevD.33.3495,2012arXiv1209.3339F}. We argue that the effects of DM dynamical friction (i.e., the drag force due to the gravitational interaction of the orbiting bodies with their wakes~\cite{Chandrasekhar:1943ys,RevModPhys.21.383,BinneyTremaine}) on the binary motion can produce observable effects and can therefore be used to put stringent constraints on the DM density in our galaxy.

\section{DM effects on the two-body problem}

We consider a binary system with masses $m_1$ and $m_2$ whose center of mass moves through a DM distribution with a roughly constant local density, $\rho_{\DM}$. In the center-of-mass frame, and neglecting relativistic effects for simplicity, the equations of motion for the two-body system read
\begin{eqnarray}
 m_i \ddot{\mathbf{r}}_i &=& \pm \frac{G m_1 m_2}{r^3}\mathbf{r}+\mathbf{F}^\DF_i , \label{eq1}
\end{eqnarray}
where the upper (lower) sign refers to $i=1$ ($i=2$), $\mathbf{r}_i$ is the position vector of the mass $m_i$, $\mathbf{r}:= \mathbf{r}_2-\mathbf{r}_1$ is the relative position of the bodies, and $\mathbf{F}^\DF_i$ is the gravitational drag on the $i$-th object. The effects of DM accretion are some orders of magnitude smaller than the effect discussed here and will be neglected.

Dynamical friction depends on the nature of the medium and on its gravitational interaction with the objects. In the case of DM, the medium can be modeled as a collisionless gas as long as the DM mean free path is much larger than the size $R_i$ of the objects. This assumption requires
\begin{equation}
 \frac{\sigma_{\DM}}{m_{\DM}} \ll 800 \left(\frac{R_\odot}{R_i}\right)\left(\frac{10^{10}{\rm GeV}/{\rm cm}^3}{\rho_{\DM}}\right) {\rm cm}^2/{\rm g}, \label{sigmaDM}
\end{equation}
where $\sigma_{\DM}$ and $m_\DM$ are the DM self-interaction cross section and the mass of the DM particles.
DM self-interactions are constrained by several observations, a conservative bound being $\sigma_{\DM}/m_{\DM}\lesssim {\rm cm}^2/{\rm g}$ (cf. Ref.~\cite{Randall:2007ph} and references therein). Therefore, even with the extreme values adopted in Eq.~\eqref{sigmaDM}, DM is perfectly collisionless.

\begin{figure}[t]
\begin{center}
\epsfig{file=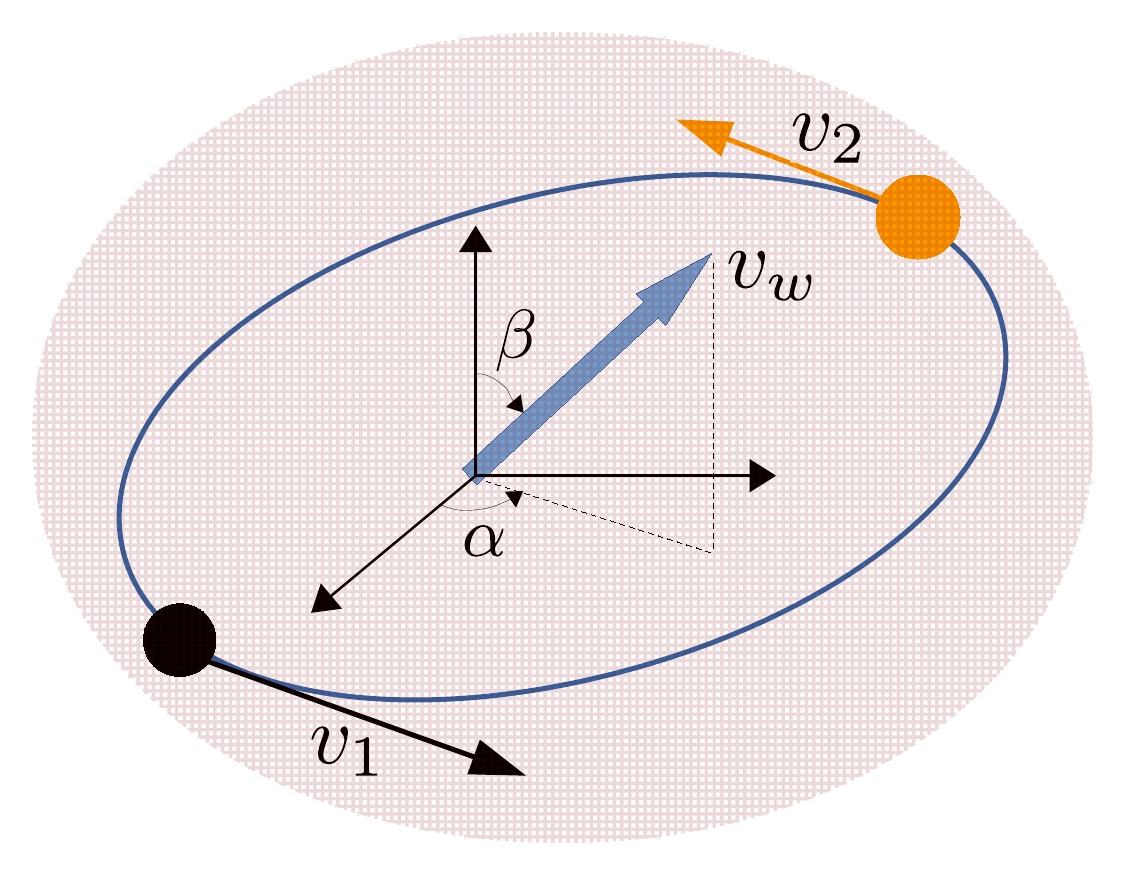,width=0.42\textwidth,angle=0,clip=true}
\caption{An illustration of the binary system under consideration in the Galactic reference frame. The center of mass of the binary moves through a constant-density DM distribution (hatched area) with velocity $\mathbf{v}_w$, experiencing a wind of DM particles with velocity $-\mathbf{v}_w$.
\label{fig:illustration}}
\end{center}
\end{figure}

For a single object in linear motion through a homogeneous collisionless medium, dynamical friction is governed by Chandrasekhar's formula~\cite{Chandrasekhar:1943ys,RevModPhys.21.383,BinneyTremaine}
\begin{equation}
 \mathbf{F}^\DF_i=-4\pi \rho_{\DM}  \lambda \frac{G^2 m_i^2 }{\tilde v_i^3}\left(\erf{(x_i)}-\frac{2x_i}{\sqrt{\pi}}e^{-x_i^2}\right)\tilde{\mathbf{v}}_i,\label{DF}
\end{equation}
where $\tilde{\mathbf{v}}_i=\dot{\mathbf{r}}_i+\mathbf{v}_w$ is the velocity of the $i$-th body relative to the DM wind, $x_i:= \tilde{v}_i/(\sqrt{2}\sigma)$,
$\sigma$ is the dispersion of the DM Maxwellian velocity distribution, and $\lambda\approx 20^{+10}_{-10}$ is the Coulomb logarithm~\cite{RevModPhys.21.383,BinneyTremaine}. We have also introduced the rotational velocity of the binary around the galaxy, $\mathbf{v}_w = v_w(\cos\alpha\sin\beta,\sin\alpha\sin\beta,\cos\beta)$, which (neglecting the rotational velocity of the DM halo) is opposite to the velocity of the wind of DM particles relative to the center of mass (cf. Fig.~\ref{fig:illustration} for an illustration of the system under consideration). Over the time scale of observations (at most some decades) $\mathbf{v}_w$ is approximately constant.

Although Eq.~\eqref{DF} was derived for linear motion, it yields remarkably precise results also for the case of a generic orbit~\cite{1983ApJ...264..364L,Kim:2007zb,2011MNRAS.411..653J}. When applied to a binary system, Chandrasekhar's formula neglects the interaction of one component with its companion's wake. This effect is negligible when the characteristic size $L$ of the wake is smaller than the orbital radius $r_0$ and when the orbital velocity $v$ is sufficiently small as to give time to the wake to disperse after an orbit. By estimating $L$ from the size of the gravitational sphere of influence, $m\sigma^2\sim G m_i m_\DM/L$, the first condition $L\ll r_0$ reads~\cite{1990ApJ...359..427B}
\begin{equation}
 P_b\gg \frac{G m_i}{\sigma^3}\sim 0.6\left(\frac{m_i}{1.3 \,M_\odot}\right)\left(\frac{150\,{\rm km/s}}{\sigma}\right)^3\,{\rm day}\,,~\label{condPb}
\end{equation}
where $P_b$ is the orbital period of the binary. This condition also implies $v\ll\sigma$, which guarantees that the particles in the wake of one object disperse before the arrival of the companion along the orbit. Thus, at least for binaries at large orbital distance, the superposition of Eq.~\eqref{DF} for $i=1,2$ is justified. For the most relativistic pulsar binaries known to date, $P_b\sim 0.1\,{\rm day}$ so the mutual interaction of the wake should provide a small effect for $\sigma\gtrsim200\,{\rm km/s}$. With these motivations, here we restrict to the condition~\eqref{condPb} and adopt Eq.~\eqref{DF} to model the gravitational drag. A more general analysis valid for any orbital period is underway. Some limitations and the range of validity of Eq.~\eqref{DF} in binary systems are discussed in Appendix~\ref{app:DF}.

\subsection{Perturbed orbits}
By introducing the center-of-mass position vector $\mathbf{R}:= (m_1 \mathbf{r}_1+m_2 \mathbf{r}_2)/M$ (with $M=m_1+m_2$), the system~\eqref{eq1} ($i=1,2$) can be conveniently written as a set of two equations of motions for $\mathbf{r}$ and $\mathbf{R}$. 
By rewriting Eq.~\eqref{DF} as $\mathbf{F}^\DF_i=-A b_i \frac{m_i^2}{M} \tilde{\mathbf{v}}_i$, where $A:= 4\pi \rho_\DM \lambda G^2 M$ and 
\begin{equation}
 b_i := \frac{1}{\tilde{v}_i^3}\left[\erf{(x_i)}-\frac{2x_i}{\sqrt{\pi}}e^{-x_i^2}\right]\,, \label{bi}
\end{equation}
it is straightforward to show that the equations of motion reduce to
\begin{eqnarray}
 \dot{\mathbf{v}}&=&-\frac{GM}{r^3}\mathbf{r}+a_1\eta\mathbf{v}+a_2 \left(\mathbf{v}_w+\mathbf{V}\right)\,, \label{E1}\\
 \dot{\mathbf{V}}&=& a_2\eta\mathbf{v}+a_3 \left(\mathbf{v}_w+\mathbf{V}\right)\,, \label{E2}
\end{eqnarray}
where we have defined $\mathbf{v}=\dot{\mathbf{r}}$, $\mathbf{V}=\dot{\mathbf{R}}$,
\begin{eqnarray}
 a_1 &=& -A (b_1+b_2)\,, \\
 a_2 &=& \frac{A}{2}\left[b_1\Delta_++b_2\Delta_-\right]\,, \\
 a_3 &=& -\frac{A}{4}\left[b_1\Delta_+^2+b_2\Delta_-^2\right]\,, 
\end{eqnarray}
$\Delta_\pm=\Delta\pm1$, $\Delta=\sqrt{1-4\eta}$, $\eta=\mu/M$, and $\mu=m_1 m_2/M$ is the reduced mass. Note that, because of the external force $\mathbf{F}^\DF_i$, the center of mass is accelerated, $\ddot{\mathbf{R}}\neq0$.

The dynamics can be greatly simplified by treating the contributions of dynamical friction perturbatively. This assumption is perfectly justified by the tiny magnitude of these corrections relative to the Keplerian terms. To compute the first-order (in the DM density) corrections, we adopt the formalism of the osculating orbits (cf. Ref.~\cite{PoissonWill} for a modern treatment). For simplicity, we focus on the corrections to a circular binary; the elliptical case is a straightforward generalization.

To zeroth order, the motion is planar and it is convenient to adopt polar coordinates where $\mathbf{r}=r_0(t)(1,\varphi_0(t),0)$ and $\Omega_0=\dot{\varphi}_0$ is the orbital angular velocity. Obviously, ${\mathbf R}$ is constant and the motion is Keplerian, $r_0 \Omega_0=(GM \Omega_0)^{1/3}=v$.

To compute the first-order corrections, all quantities that are multiplied by $\rho_\DM$ can be evaluated using the Keplerian solution. In particular, to zeroth order $\mathbf{V}=\mathbf{V}_0=0$ and the right-hand side of Eqs.~\eqref{E1} and \eqref{E2} can be further simplified. To first order, Eq.~\eqref{E1} has the form $\dot{\mathbf{v}}=-\frac{GM }{r^3}\mathbf{r}+\mathbf{f}$, with $\mathbf{f}=a_1\eta\mathbf{v}+a_2\mathbf{v}_w$, and does not depend on the motion of the center of mass.
Following Ref.~\cite{PoissonWill}, the adiabatic-evolution equations for the orbital parameters in the quasicircular case are
\begin{eqnarray}
 \dot a&=& 2\sqrt{\frac{r_0^3}{GM}} {\cal S}(t) \,, \label{dadt}\\
 \dot e&=& \sqrt{\frac{r_0}{GM}}\left[{\cal R}(t)\sin\Omega_0t +2{\cal S}(t)\cos\Omega_0t \right]\,, \label{dedt}\\
 \dot\iota&=& \sqrt{\frac{r_0}{GM}} {\cal W}(t)\cos(\Omega_0 t+\omega)\,, \label{didt}\\
 \dot\Omega&=& \frac{1}{\sin \iota} \sqrt{\frac{r_0}{GM}}  {\cal W}(t) \sin(\Omega_0 t+\omega)\,, \label{dOmegadt}
\end{eqnarray}
where $a$, $e$, $\omega$, $\iota$ and $\Omega$ are the semiaxis major, the eccentricity, the longitude of pericenter, the inclination, and the longitude of the ascending node, respectively. The source terms in the equations above read
\begin{eqnarray}
 {\cal R}&:=& \mathbf{f}\cdot \mathbf{n} = a_2 \,\mathbf{n} \cdot \mathbf{v}_w =a_2 v_w \sin\beta \cos(\Omega_0 t-\alpha)\,,\\
 {\cal S}&:=& \mathbf{f}\cdot \bm{\lambda} = a_1\eta \,\bm{\lambda}\cdot \mathbf{v}+a_2\, \bm{\lambda}\cdot \mathbf{v}_w \nn\\
 &=& a_1\eta v-a_2 v_w\sin\beta\sin(\Omega_0t-\alpha)\,,\\
 {\cal W}&:=& \mathbf{f}\cdot \mathbf{e}_z = a_2 v_w\cos\beta\,, \label{eqW}
\end{eqnarray}
where we used the definitions of the orbital basis vectors to zeroth order, namely $\mathbf{n}=(\cos\Omega_0t,\sin\Omega_0t,0)$, $\bm{\lambda}=(-\sin\Omega_0t,\cos\Omega_0t,0)$, $\mathbf{e}_z=(0,0,1)$. As expected by the fact that the Keplerian motion is periodic, the source terms display a peculiar modulation with period equal to the orbital period $P_b:=2\pi/\Omega_0$.

A relevant quantity is the time derivative of the orbital period which --~in a quasiadiabatic approximation valid in the perturbative regime~-- can be obtained from Eq.~\eqref{dadt} and from Kepler's third law, $\dot P_b= \frac{3 P_b}{2r_0}\dot a$.
Using Eq.~\eqref{dadt}, we obtain
\begin{equation}
 \dot P_b^\DF(t) = \frac{3 v P_b}{2}\left[a_1\eta -a_2 \Gamma\sin\beta\sin(\Omega_0t-\alpha)\right]\,, \label{PBDOTDF}
\end{equation}
where $\Gamma:=v_w/v$. We focus on the secular changes of the orbital parameters, so henceforth we shall consider orbital-averaged quantities obtained from Eqs.~\eqref{dadt}--\eqref{dOmegadt}, e.g. $\langle X\rangle:= P_b^{-1}\int_0^{P_b} dt  X(t)$.


\subsection{Other contributions to $\dot P_b$}
In addition to the changes of the orbital period induced by the secular evolution of the semiaxis major [cf. Eq.~\eqref{PBDOTDF}], the observed variation of $P_b$ in a binary system is affected by kinematic effects, namely by the center-of-mass acceleration along the line of sight and by the variation of the orbital inclination $\iota$. Both effects produce a Doppler shift of pulse frequencies which affects the measurement of $P_b$ (cf., e.g., Ref.~\cite{HandbookPulsar}).

The induced change in the orbital period due to the center-of-mass acceleration reads
\begin{equation}
 \dot P_b^{\rm cm} = P_b \dot{\mathbf{V}}\cdot \mathbf{e}_Z\,, \label{CM}
\end{equation}
where $\mathbf{e}_Z$ is the unit vector parallel to the line of sight as defined in Ref.~\cite{PoissonWill}. In Appendix~\ref{app:CM} we compute the contribution above, whose final expression reads
\begin{eqnarray}
 \dot P_b^{\rm cm}(t)&&= v P_b \left\{a_2\eta  \sin\iota\cos(\Omega_0t+\omega)\right.\nn\\
 &&\left.+a_3 \Gamma\left[\cos\iota\cos\beta+\sin\iota\sin\beta\sin(\alpha+\omega)\right]\right\}. \label{PBDOTcm}
\end{eqnarray}

Finally, the contribution of the variation of the orbital inclination reads $\dot P_b^\iota = \frac{3}{2} \tan\iota P_b \dot\iota$ which, by using Eqs.~\eqref{didt} and \eqref{eqW}, reduces to
\begin{equation}
 \dot P_b^\iota(t) = \frac{3}{2}a_2 \Gamma P_b \tan\iota \cos\beta \cos(\Omega_0t+\omega)\,. \label{PBDOTiota}
\end{equation}
Note that $\dot P_b^\iota(t)=0$ when $\iota=0$ (i.e. when the orbit is perpendicular to the line of sight) and when $\beta=\pi/2$ (i.e. when the DM wind blows parallel to the orbital plane, cf. Fig.~\ref{fig:illustration}).

The total induced change in the orbital period is $\dot P_b^\DF+\dot P_b^{\rm cm}+\dot P_b^\iota$, obtained from Eqs.~\eqref{PBDOTDF}, \eqref{PBDOTcm} and \eqref{PBDOTiota}. In our case the contribution of $\dot P_b^{\rm cm}$ is typically subdominant and can be neglected, whereas $\dot P_b^\iota$ can be of the same order of $\dot P_b^\DF$.

\subsection{Analytical results}\label{sec:ana}
In generic situations the orbital average of Eqs.~\eqref{PBDOTDF}, \eqref{PBDOTcm}, and \eqref{PBDOTiota} should be performed numerically because  the quantity $b_i(t)$ defined in Eq.~\eqref{bi} (and evaluated using the Keplerian solution) is a cumbersome function of the time. Before presenting the results of such numerical integration in Sec.~\ref{sec:num}, here we discuss some limits in which the secular changes can be computed analytically.
For simplicity we focus on the $\dot P_b^\DF$ term; the corrections $\dot P_b^{\rm cm}$ and $\dot P_b^\iota$ can be computed using the same procedure.

\subsubsection{Large-$\sigma$ limit} 
In the limit $\sigma\gg v_w,v$, from Eq.~\eqref{bi} we obtain
\begin{eqnarray}
 b_i&&\sim \sqrt{\frac{2}{9\pi}}\frac{1}{\sigma^3}\nn\\
 &&\times\left\{1-\frac{3v^2}{40\sigma^2}\left[4 \Gamma^2+\Delta_\mp^2-4 \Gamma\Delta_\mp\sin\beta\sin(\Omega_0t-\alpha)\right]+\dots\right\}\,,\nn
\end{eqnarray}
where we have used the Keplerian solution to lowest order, and the upper (lower) sign refers to $i=1$ ($i=2$). Note that, to leading order, $b_1\sim b_2\sim{\rm const}$. By inserting the above expression into Eq.~\eqref{dadt}, it is straightforward to obtain the analytical result
\begin{widetext}
 \begin{eqnarray}
 \dot{P}^\DF_{b}(t) &\sim& -8 \sqrt{2\pi} G^2 \frac{\mu \lambda \rho_\DM P_b}{\sigma^3}\left\{1+\frac{\Gamma \Delta}{2\eta}\sin\beta\sin(\Omega_0t-\alpha)+\right.\nn\\
 &&\left.+\frac{3v^2}{20\eta\sigma^2}\left[\Gamma ^2 \eta  \cos(2 \Omega_0t-2 \alpha)+\Gamma 
   \Delta  \left(3 \eta -\Gamma ^2\right) \sin\beta
   \sin(\Omega_0t-\alpha)+\eta  \left(2\eta-1-3 \Gamma ^2 +2 \Gamma ^2 \cos2 \beta  \sin^2(\Omega_0t-\alpha)\right)\right]\right\}\,,  \nn
\end{eqnarray}
\end{widetext}
where the terms on the last line are the next-to-leading-order corrections in the large-$\sigma$ limit.
The orbital change as a function of time has a characteristic modulation with period $P_b$, an amplitude proportional to the DM wind velocity, and a phase given by the angle $\alpha$. This modulation is expected, due to the periodic motion at Keplerian order. However, to leading order the oscillatory term averages to zero over an orbital period. In this limit the secular decay,
\begin{eqnarray}
 \langle\dot{P}^\DF_{b}\rangle &&\sim -8 \sqrt{2\pi} G^2 \frac{\mu \lambda \rho_\DM P_b}{\sigma^3}\nn\\
 &&\times\left[1+\frac{3v^2}{20\sigma^2}(2\eta-1-3\Gamma^2+\Gamma^2\cos2\beta)\right]\,, \label{PdotDF}
\end{eqnarray}
does not depend on the DM wind $v_w$ to leading order. Equation~\eqref{PdotDF} extends the result by Gould~\cite{1991ApJ...379..280G} obtained in the large-$\sigma$ regime within a rather different framework; in fact, the leading-order term of the expression above matches exactly that derived in Ref.~\cite{1991ApJ...379..280G}. This agreement gives further confirmation of the validity of Eq.~\eqref{DF}, at least in the $\sigma\gg v$ limit.

Note also that, to leading order, $a_2$ in Eq.~\eqref{PBDOTiota} is a constant and therefore $\langle \dot P_b^\iota\rangle=0$. Likewise, when $\beta=0$ the orbital average $\langle \dot P_b^\iota\rangle=0$ in general (and not only in the large-$\sigma$ limit), because in such case $a_2$ does not depend on time.
\subsubsection{Large-$v_w$ limit}  
Another relevant limit is $v_w\gg \sigma,v$, which yields
\begin{eqnarray}
 b_i&&\sim \frac{1}{v_w^3}\left[1+\frac{3\Delta_\mp}{2\Gamma}\sin\beta\sin(\Omega_0 t-\alpha)\right]\,,
\end{eqnarray}
where again the upper (lower) sign refers to $i=1$ ($i=2$). From this expression it is straightforward to obtain
\begin{widetext}
\begin{eqnarray}
 \dot{P}^\DF_{b}(t) &\sim& -\frac{24 G^2 M \pi ^2 \Delta  \lambda  \rho_\DM}{v v_w^2
   \Omega_0}\left[\sin\beta \sin(\Omega_0t -\alpha)+\frac{2 \eta  \left[1-3
   \sin^2\beta \sin^2(\Omega_0 t-\alpha)\right]}{\Gamma \Delta }+\frac{3 \eta  \sin\beta \sin(\Omega_0t -\alpha)}{\Gamma^2}\right]\,.\nn
\end{eqnarray}
\end{widetext}
The ${\cal O}(1/v_w^2)$ and ${\cal O}(1/v_w^4)$ terms are zero when $\beta=0$ and their orbital average vanishes for any $\beta$. Therefore, in this limit the secular change of the orbital period reads
\begin{eqnarray}
 \langle\dot{P}^\DF_{b}\rangle &\sim& -\frac{6 \pi G^2 M P_b  \eta  \lambda  \rho_\DM (1+3\cos2\beta)}{v_w^3}\,. \label{PdotDF2}
\end{eqnarray}
The leading-order coefficient depends on the direction of the DM wind but not on the DM velocity dispersion.

Interestingly, $\langle\dot{P}^\DF_{b}\rangle$ becomes positive when $55^{\circ}\lesssim\beta\lesssim 125^{\circ}$. In this case the orbit tends to outspiral (i.e., to get softer~\cite{Quinlan:1996vp}) due to the effect of DM dynamical friction.
This configuration seems to violate Heggie's law~\cite{1975MNRAS.173..729H} according to which hard binaries (i.e. those with binding energy $E_b\gg m_\DM\sigma^2$) get harder ($\langle\dot{P}^\DF_{b}\rangle<0$) in a medium of fast low-mass objects. This result might have implications in studies of Galactic dynamics and we plan to investigate it more in detail in a separate publication. Our result is not in contrast with the findings of Ref.~\cite{1991ApJ...379..280G}, because the latter were obtained for $v_w=0$, whereas here we assume $v_w\gg\sigma,v$. 
Note that, in addition to the large-$v_w$ limit, the condition~\eqref{condPb} also needs to be satisfied.
%

\subsubsection{Small-$v_w$ limit}
In the limit $v_w\sim 0$, we obtain
\begin{eqnarray}
 \dot{P}^\DF_{b}(t) &\sim& -\frac{6 G M  \mu \lambda    \rho_\DM P_b^2}{\sqrt{\pi}m_1^3 m_2^3 \sigma}\nn\\
 &\times&\left\{M \sqrt{\pi } \left[m_2^3  \erf(y_1)+ m_1^3  \erf(y_2)\right]\sigma  \right.\nn\\
 &&\left.-\sqrt{2} m_1 m_2 \left(e^{-y_2^2} m_1^2+e^{-y_1^2} m_2^2\right) v\right\}\,. \label{PdotDF3}
\end{eqnarray}
where $y_i:=\frac{m_i v}{\sqrt{2} M\sigma }$. Since in this case $\dot{P}^\DF_{b}$ does not depend on time, the orbital average coincides with the equation above, which also reduces to the leading-order term of Eq.~\eqref{PdotDF} when $\sigma\gg v$.
Note that in this case $\langle\dot{P}^\DF_{b}\rangle<0$, in agreement with the results of Ref.~\cite{1991ApJ...379..280G} and with Heggie's law.
%

\subsection{Numerical results}\label{sec:num}
%
\begin{figure}[th]
\begin{center}
\epsfig{file=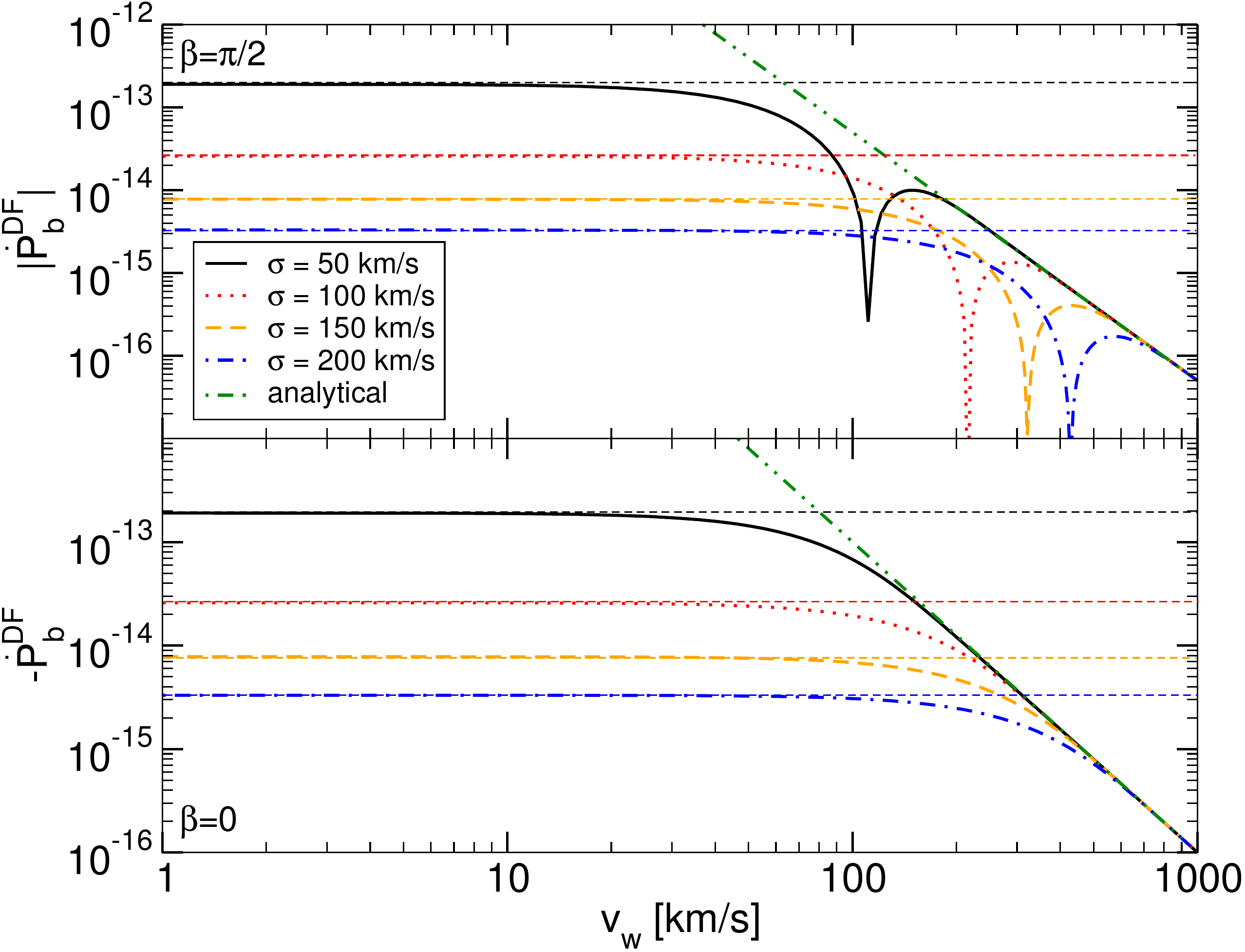,width=0.47\textwidth,angle=0,clip=true}
\caption{Secular change of the orbital period of a binary pulsar due to DM dynamical friction as a function of the DM wind velocity $v_w$, for different values of the DM velocity dispersion, and for $\beta=\pi/2$ (top panel) and $\beta=0$ (bottom panel). We assumed the following reference values: $P_b=100\,{\rm day}$, $m_1=1.3 M_\odot$, $m_2=0.3 M_\odot$ and $\rho_\DM=2\times 10^3\,{\rm GeV/cm}^3$. The analytical result refers to Eq.~\eqref{PdotDF2}, whereas the horizontal lines refer to Eq.~\eqref{PdotDF3}.
\label{fig:PBDOT}}
\end{center}
\end{figure}

In various situations of phenomenological interest $v\sim v_w\sim\sigma\sim{\cal O}(100\,{\rm km/s})$ and the analytical limits discussed above are not valid. In this case one needs to resort to a numerical integration, which is in any case straightforward given the simple form of Eqs.~\eqref{dadt}--\eqref{dOmegadt}. Henceforth the notation $\langle X \rangle$ for the orbital average of a quantity $X(t)$ will be omitted.

In Fig.~\ref{fig:PBDOT} we present some exact numerical results for $\dot P_b^\DF$ and compare them with Eqs.~\eqref{PdotDF2} and~\eqref{PdotDF3}. 
As expected, Eq.~\eqref{PdotDF2} is an excellent approximation to the exact result when $v_w\gg \sigma,v$. Furthermore, we find that the analytical result~\eqref{PdotDF3} is accurate in the entire range $v_w\lesssim \sigma$ and not only when $v_w\to0$, at least for long orbital periods.
Note also that $\dot P_b$ can become positive above a critical value of $v_w$ when $\beta$ is sufficiently large. This is in agreement with Eqs.~\eqref{PdotDF3} and \eqref{PdotDF2}. While the former equation predicts $\dot P_b<0$ when $v_w\sim0$, the latter predicts $\dot P_b>0$ when $55^{\circ}\lesssim\beta\lesssim 125^{\circ}$ and at large values of $v_w$. By continuity, there exists a value of $v_w$ at which $\dot P_b$ changes sign.

\begin{figure*}[th]
\begin{center}
\epsfig{file=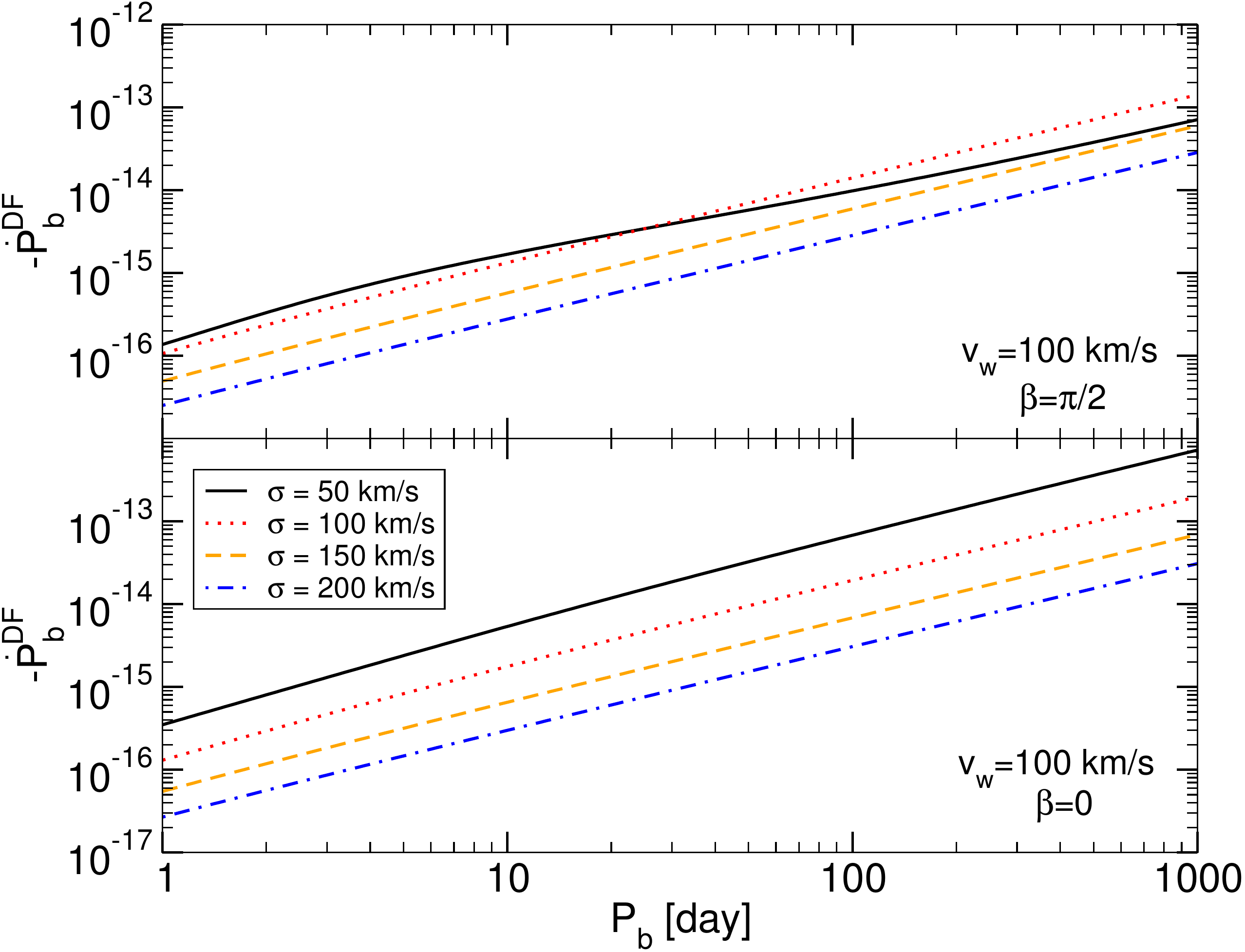,width=0.47\textwidth,angle=0,clip=true} 
\epsfig{file=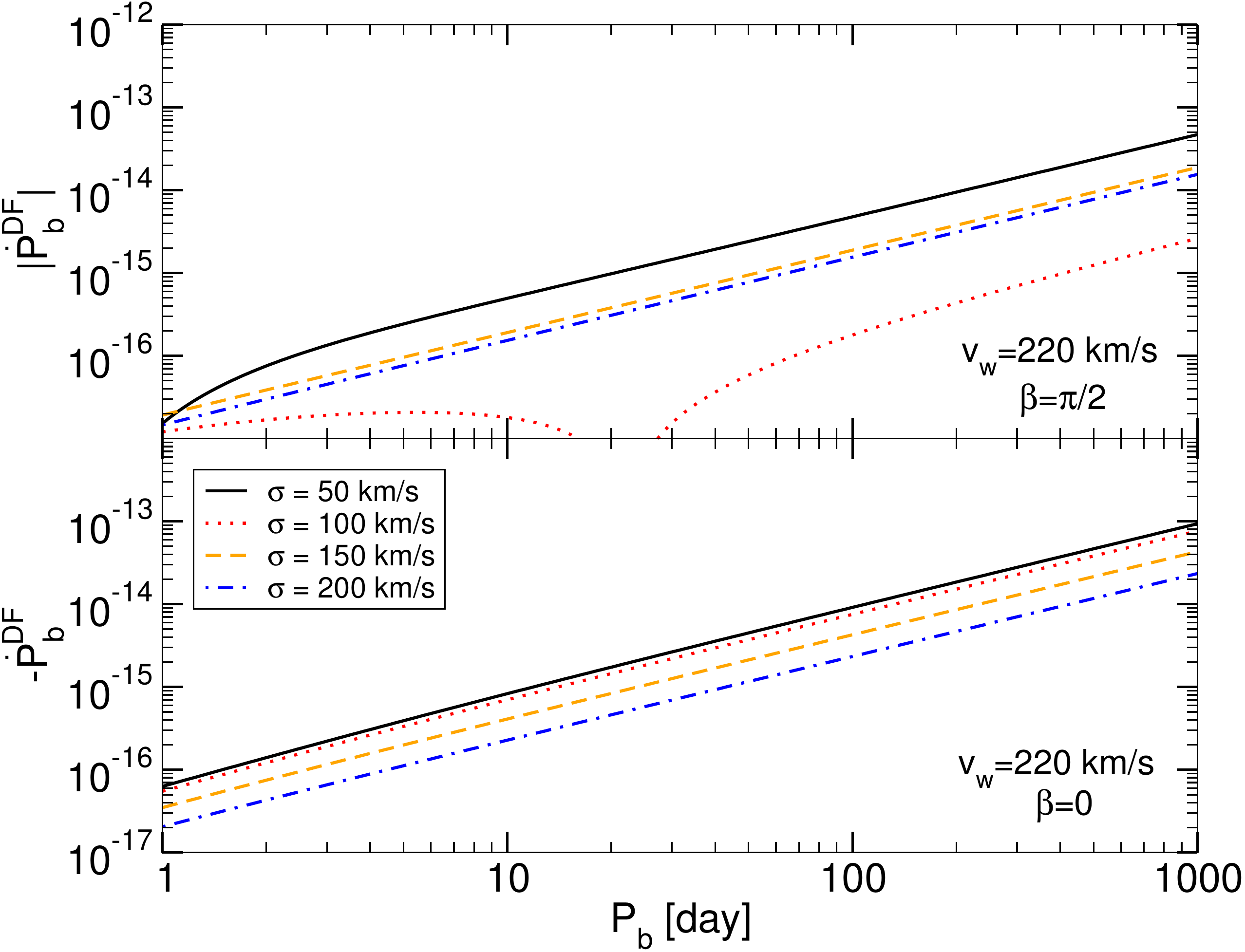,width=0.47\textwidth,angle=0,clip=true} 
\caption{Secular change of the orbital period of a binary pulsar due to DM dynamical friction as a function of the orbital period $P_b$, for different values of the DM velocity dispersion and of the DM wind (left panels: $v_w=100\,{\rm km/s}$; right panels: $v_w=220\,{\rm km/s}$), and for $\beta=\pi/2$ (top panels) and $\beta=0$ (bottom panels). We considered the same reference values as in Fig.~\ref{fig:PBDOT}.
\label{fig:PBDOT2}}
\end{center}
\end{figure*}

The secular change $\dot P_b^\DF$ as a function of the orbital period $P_b$ is shown in Fig.~\ref{fig:PBDOT2} for various choices of the angle $\beta$, of the DM wind velocity $v_w$, and of the DM velocity dispersion $\sigma$. As expected from Eq.~\eqref{DF}, corrections are larger for longer orbital periods, although in some particular cases (when $\beta\sim\pi/2$) the behavior of the function $\dot P_b$ might be nonmonotonic\footnote{As previously discussed, $\dot P_b$ can change sign for large values of $\beta$. In the logarithmic scale adopted for the vertical axis of Fig.~\ref{fig:PBDOT2} (and of subsequent figures), the zero crossing corresponds to ``cusps'' which occasionally appear in the function $\dot P_b^\DF$ when $\beta\approx \pi/2$.}

\begin{figure*}[t]
\begin{center}
\epsfig{file=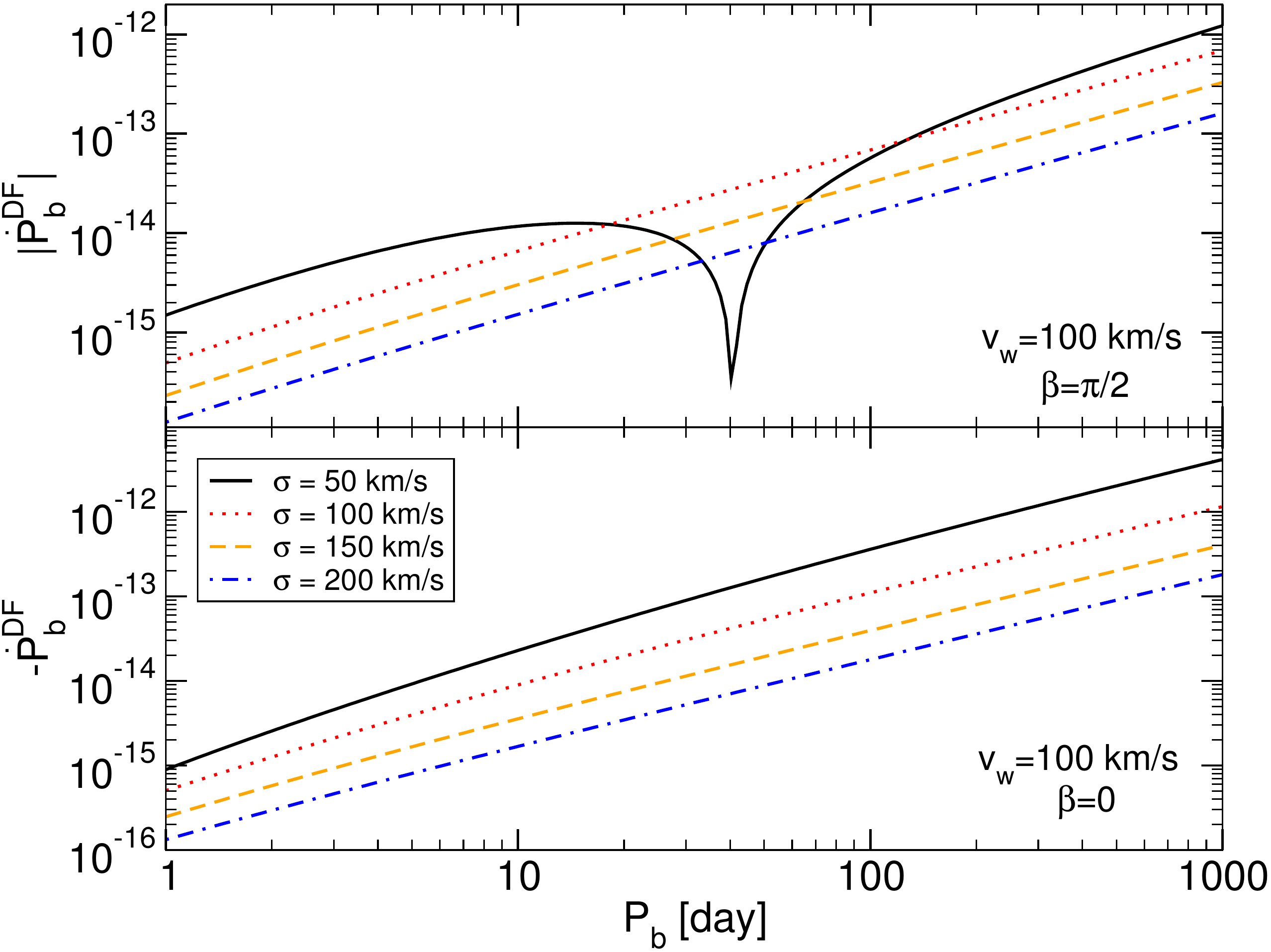,width=0.47\textwidth,angle=0,clip=true} 
\epsfig{file=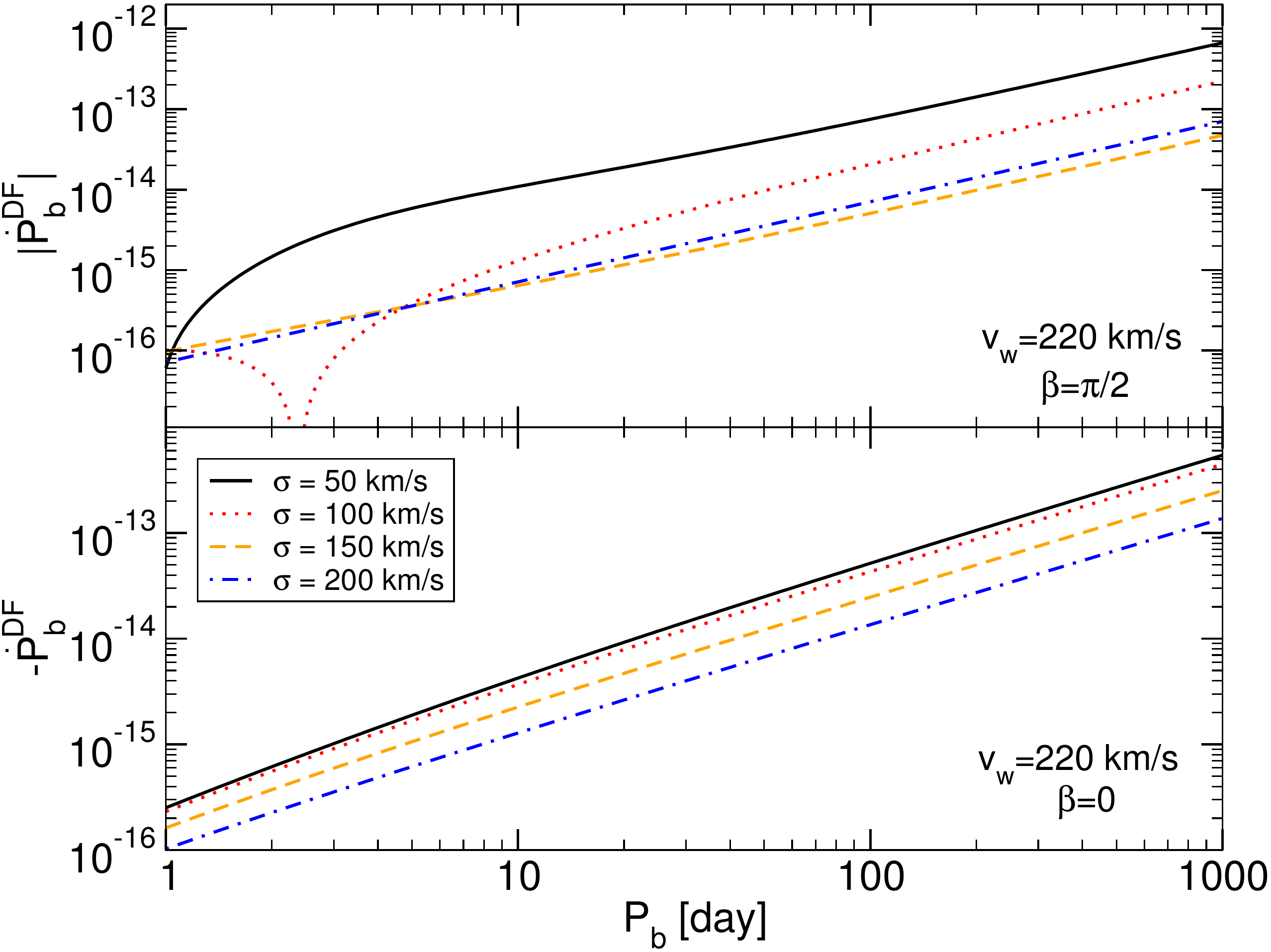,width=0.47\textwidth,angle=0,clip=true} 
\caption{Same as in Fig.~\ref{fig:PBDOT2} but for $m_1=2 M_\odot$ and $m_2=5 M_\odot$.
\label{fig:PBDOT3}}
\end{center}
\end{figure*}

An extensive numerical search shows that the secular change $\dot P_b^\DF$ is independent of $\alpha$ (in agreement with the analytical results previously presented) and that the corrections depend only mildly on the value of $\beta$. In Figs.~\ref{fig:PBDOT} and~\ref{fig:PBDOT2} we consider a typical neutron star-white dwarf system with $m_1\approx 1.3 M_\odot$ and $m_2\approx 0.3 M_\odot$. This is a conservative assumption, since the secular changes grow with the masses of the binary, as shown in Fig.~\ref{fig:PBDOT3} (to be compared with the corresponding Fig.~\ref{fig:PBDOT2}).

Finally, our perturbative solution is valid for $t\ll P_b/\dot P^\DF_b$, i.e. for $t\ll 10^{12}P_b$ or longer (cf. Figs.~\ref{fig:PBDOT2} and~\ref{fig:PBDOT3}) and it is therefore perfectly reliable during the observation time (lasting at most some decades).

%

\section{Constraints on the DM density}
The orbital period of various binary pulsars is found to be constant over several years within astonishing precision. Such systems are therefore ideal candidates to constrain the small effect of DM on the binary motion. For some systems, measurements of the small secular changes of the orbital period ($\dot P_b$, $\ddot P_b$, ...) are also available. 

The apparent orbital change $\dot P^\obs_b$ needs to be corrected to take into account the differential acceleration in the Galactic gravitational potential~\cite{1991ApJ...366..501D} and the transverse motion of the binary relative to Earth (Shklovskii's effect~\cite{1970SvA....13..562S}). After subtracting these kinematic contributions, the intrinsic orbital decay, $\dot P_b^\INT$, can still be caused by gravitational-wave emission, mass loss, tidal torques and gravitational quadrupole coupling~\cite{Doroshenko:2001zi,Lazaridis:2011wc,Pletsch:2015nia} (cf. Ref.~\cite{HandbookPulsar} for an overview).

For some binary systems these corrections can be computed precisely. For relativistic binaries, the dominant contribution is typically due to the gravitational-wave emission, $\dot P_b^\GW$, and the ``excess'' orbital decay, $\dot P_b^{\rm xs}=\dot P_b^\INT-\dot P_b^\GW$, is found to be consistent with zero within errors. In other words, for some systems the emission of gravitational waves as predicted by the quadrupole formula of general relativity~\cite{Peters:1963ux} completely accounts for the observed intrinsic secular change of $\dot P_b$. This is the case of J1738+0333, whose measured excess orbital period, $|\dot P_b^{\rm xs}|\lesssim 2\times 10^{-15}$, provides one of the most stringent constraints on modified theories of gravity~\cite{2012MNRAS.423.3328F}.

Interestingly, the typical amplitude of $\dot P_b^\DF$ shown in Figs.~\ref{fig:PBDOT}--\ref{fig:PBDOT3} is comparable to or even larger than the observed value of $\dot P_b^{\rm xs}$ for some systems~\cite{Doroshenko:2001zi,Lazaridis:2011wc,Pletsch:2015nia,2012MNRAS.423.3328F,Reardon:2015kba}. As an order-of-magnitude estimate, the leading term of Eq.~\eqref{PdotDF} can be rewritten as
\begin{eqnarray}
 \dot{P}^\DF_{b} &\approx&-3\times 10^{-14} \frac{\mu_1\lambda_{20} \rho_\DM^\GC P_b^{(100)}}{\sigma_{150}^{3}}\, \label{PdotDFb}
\end{eqnarray}
where $\mu_1=\mu/M_\odot$, $\lambda_{20}= {\lambda}/{20}$, $P_b^{(100)}={P_b}/(100\,{\rm day})$, $\sigma_{150}=\sigma/(150\,{\rm km/s})$, and $\rho_\DM^\GC=\rho_\DM / (2\times 10^3\,{\rm GeV/cm}^3)$ is normalized by approximately the value of the DM density at a distance $D\approx5\,{\rm pc}$ from the Galactic center in the case of a typical Navarro-Frenk-White profile~\cite{Navarro:1995iw}. For the same profile, the DM density at $D\approx1\,{\rm pc}$ can be larger by a factor $100$ in the case of adiabatic growth of the central BH~\cite{Gondolo:1999ef,Ullio:2001fb}.
Therefore, a binary pulsar with $P_b\sim 100\,{\rm day}$ near the Galactic center will display a secular change of the orbital period well within the current accuracy of pulsar timing. 

On the other hand, Eq.~\eqref{PdotDFb} can be inverted yielding an approximate upper bound on the DM density allowed by the timing measurement of a binary pulsar with orbital period $P_b$ and excess orbital change $\dot P_b^{\rm xs}$, namely
\begin{eqnarray}
 \rho_\DM &\lesssim&10^2 \left(\frac{\dot P_b^{\rm xs}}{10^{-15}}\right) \frac{\sigma_{150}^{3}}{ \mu_1\lambda_{20}  P_b^{(100)} }\, {\rm GeV/cm}^3\,. \label{rhomax}
\end{eqnarray}

\begin{figure*}[t]
\begin{center}
\epsfig{file=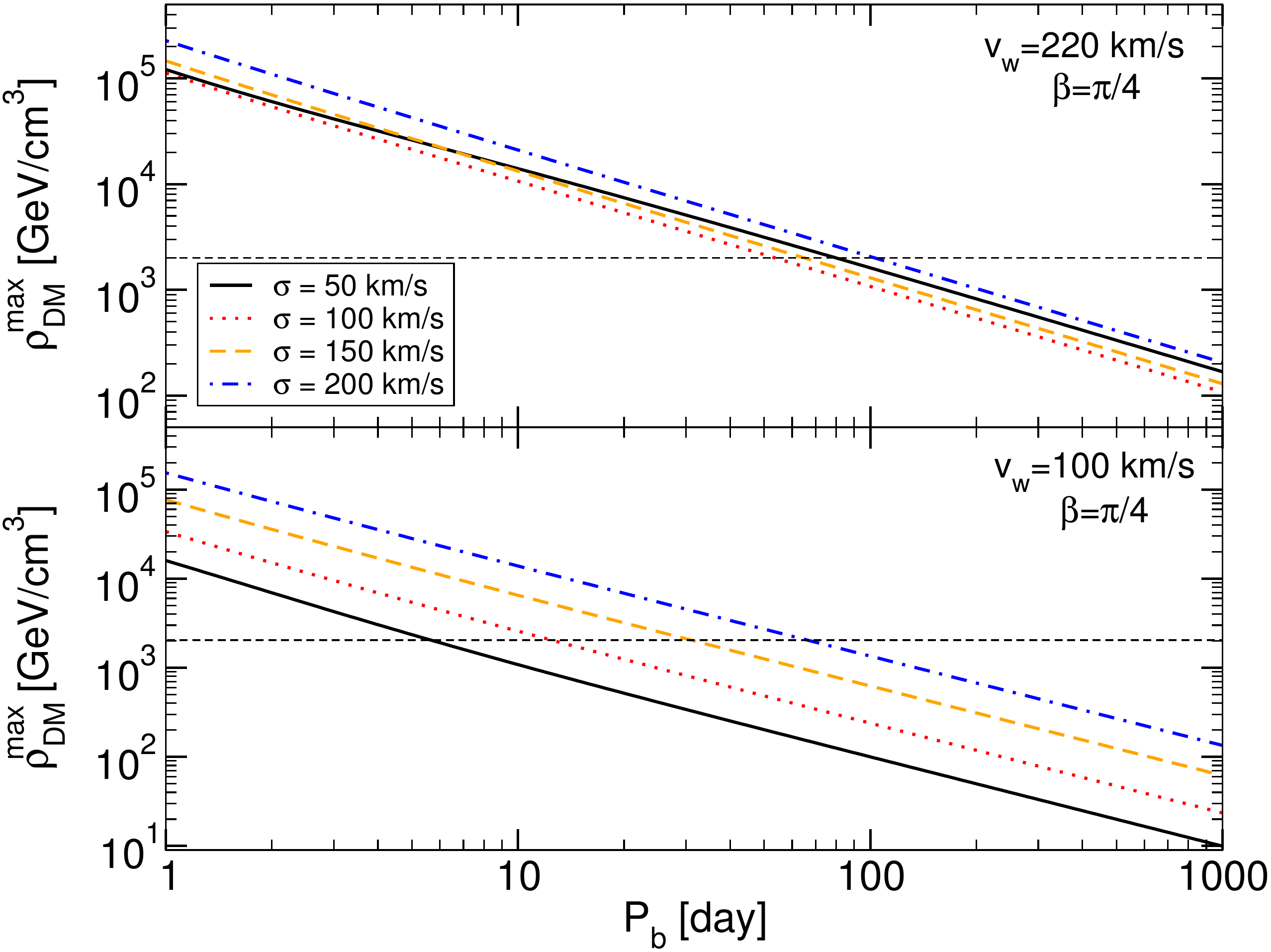,width=0.47\textwidth,angle=0,clip=true} 
\epsfig{file=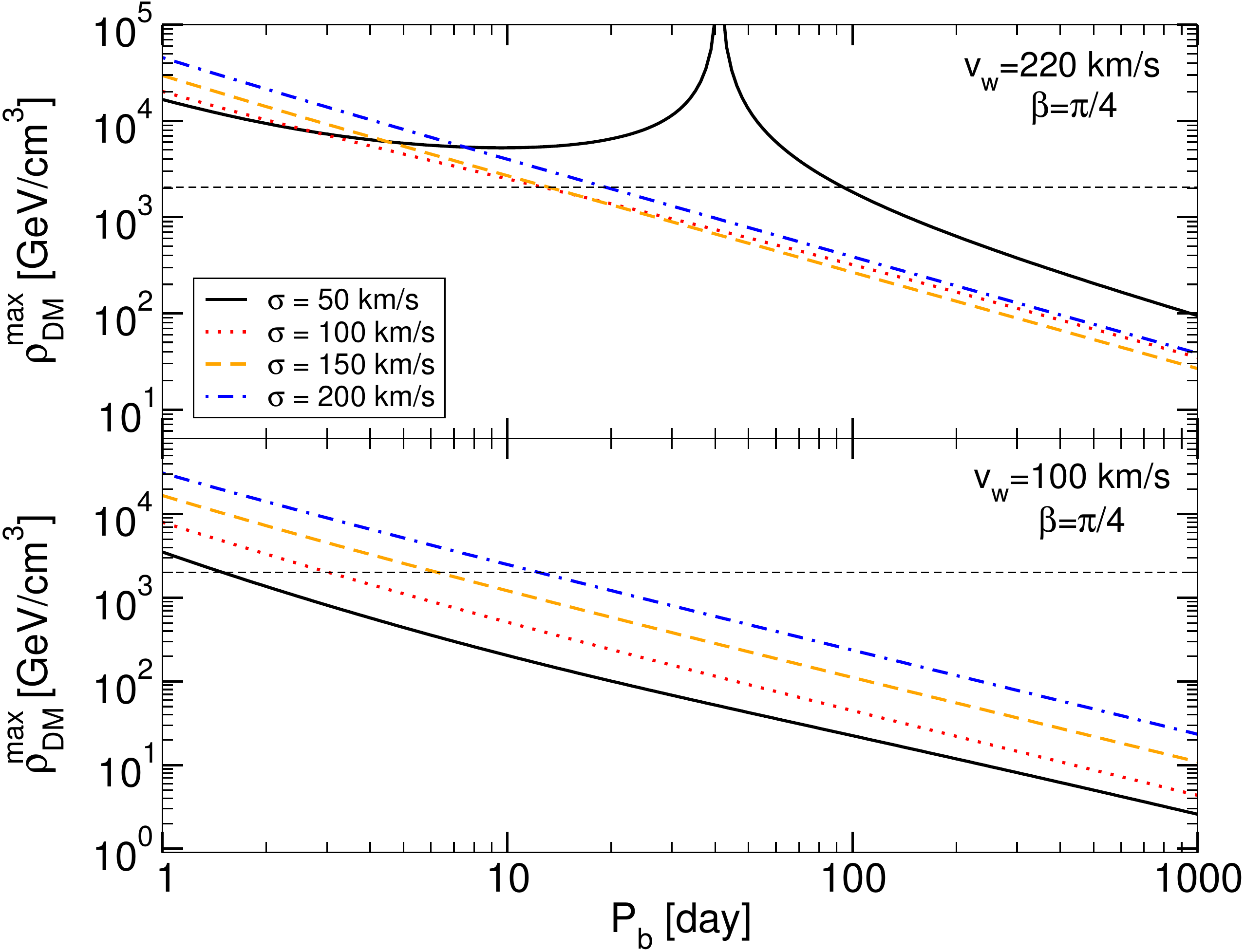,width=0.47\textwidth,angle=0,clip=true} 
\caption{Upper bounds on the DM density due to DM dynamical friction as a function of the orbital period $P_b$ of a binary pulsar with excess secular change $|\dot P_b^{\rm xs}|\lesssim2\times10^{-15}$ and for different choices of the DM velocity dispersion $\sigma$. We considered $\beta=\pi/4$ and two representative values of the DM wind velocity in the top and bottom panels, respectively. Left panels: $m_1=1.3 M_\odot$ and $m_2=0.3 M_\odot$. Right panel: $m_1=2 M_\odot$ and $m_2=5 M_\odot$. The horizontal line denotes the value $\rho_\DM\approx 2\times 10^3\,{\rm GeV/cm}^3$, corresponding to a distance $D\approx5\,{\rm pc}$ from the Galactic center in the case of a typical Navarro-Frenk-White profile~\cite{Navarro:1995iw}.
\label{fig:bounds}}
\end{center}
\end{figure*}

The bound above is purely indicative, since it is obtained using the analytical result~\eqref{PdotDF} valid only in the limit $\sigma\gg v_w,v$. Figure~\ref{fig:bounds} shows more precise upper bounds on $\rho_\DM$ derived from the numerical results by imposing $|\dot P_b^\DF|\lesssim 2\times 10^{-15}$, i.e. assuming a measurement of $\dot P_b^{\rm xs}$ as precise as that for J1738+0333~\cite{2012MNRAS.423.3328F}.
Note that the values $\sigma=150\,{\rm km/s}$, $v_w=220\,{\rm km/s}$ and $\sigma=50\,{\rm km/s}$, $v_w=100\,{\rm km/s}$ considered in Fig.~\ref{fig:bounds} are representative of the DM velocity dispersions and Galactic rotation velocities near the Solar System and close to the Galactic center, respectively.

Due to a selection bias, most binary pulsars are detected near Earth. Indeed, we searched in the ATNF Pulsar Catalogue~\cite{catalogue,Manchester:2004bp} for those binaries whose (apparent) orbital change $\dot P^\obs_b$ has been measured and found $32$ systems which  are located at a distance $D\in[1-12]\,{\rm kpc}$ from the Galactic center. 
Therefore, even though for some of these binaries the intrinsic orbital change can be as small as $\dot P_b^\INT\sim 10^{-14}$, the constraints shown in Fig.~\ref{fig:bounds} are not competitive compared to the local value of DM density, $\rho_\DM\approx 0.4\,{\rm GeV/cm}^3$.
Nonetheless, Fig.~\ref{fig:bounds} shows that detecting a binary pulsar at $D\lesssim 10\,{\rm pc}$ would give strong constraints on theoretical DM halo profiles and on the evolution of the central BH, e.g. on the formation of DM spikes~\cite{Gondolo:1999ef}.

Note that the most relativistic binary pulsars have $P_b={\cal O}(0.1)\,{\rm day}$, including J1738+0333 which is the system with the most precise measurement of $\dot P_b^{\rm xs}$ consistent to zero known to date~\cite{2012MNRAS.423.3328F}. As previously discussed, our analysis does not apply to these systems, because Eq.~\eqref{condPb} is not satisfied. On the other hand, at least in the large-$P_b$ regime, the effect of DM dynamical friction grows linearly with $P_b$ so the most stringent upper bounds come from binary systems at large orbital distance.

\subsection{Upper limit on DM density derived from the timing of J1713+0747}

Binary pulsar J1713+0747~\cite{1993ApJ...410L..91F} is particularly well suited for our analysis, due to its long orbital period, $P_b\approx 67.8\,{\rm day}$, and to its small intrinsic orbital decay, $|\dot P_b^\INT|\lesssim2\times10^{-13}$~\cite{0004-637X-809-1-41}. Since gravitational-wave dissipation is negligible for this system ($\dot P_b^\GW\approx -6\times10^{-18}$, as predicted by the quadrupole formula~\cite{Peters:1963ux}), $\dot P_b^{\rm xs}\approx \dot P_b^\INT$. 
For this system, dynamical friction can be larger than gravitational-wave dissipation by several orders of magnitude and would be the dominant contribution to the intrinsic orbital-period decay.
J1713+0747 is in a nearly circular orbit ($e\approx 8\times 10^{-5}$, so our assumption of circular motion at Keplerian order is accurate) and it is formed by a neutron star with $m_1\approx 1.31 M_\odot$ and by a white dwarf with $m_2\approx 0.29 M_\odot$. Its measured inclination and longitude of the pericenter are $\iota\approx 75.3^\circ$ and $\omega\approx176.2^\circ$, respectively~\cite{0004-637X-809-1-41}, whereas the distance from the Galactic center is about $D\approx7.2\,{\rm kpc}$.
We therefore assume $\sigma\approx 150\,{\rm km/s}$ and $v_w\sim (240\pm 31)\,{\rm km/s}$. The latter range is estimated by adding the measured transverse velocity\footnote{In principle, the values of $v_w$, $\alpha$, and $\beta$ can be extracted from a measurement of the three-dimensional velocity of the binary. However, for most binary pulsars, an estimate of the radial velocity is not available. An exception is J1738+0333 whose three-dimensional velocity has been recently estimated~\cite{2012MNRAS.423.3328F}. For our heuristic purposes, the estimate $v_w=v_\odot\pm v_T$ is sufficiently accurate.} of J1713+0747, $v_T\sim31.28\,{\rm km/s}$~\cite{catalogue,Manchester:2004bp}, to the Galactic rotational velocity at the Sun location, $v_\odot\approx 240\,{\rm km/s}$~\cite{Reid:2014boa}.

\begin{figure}[th]
\begin{center}
\epsfig{file=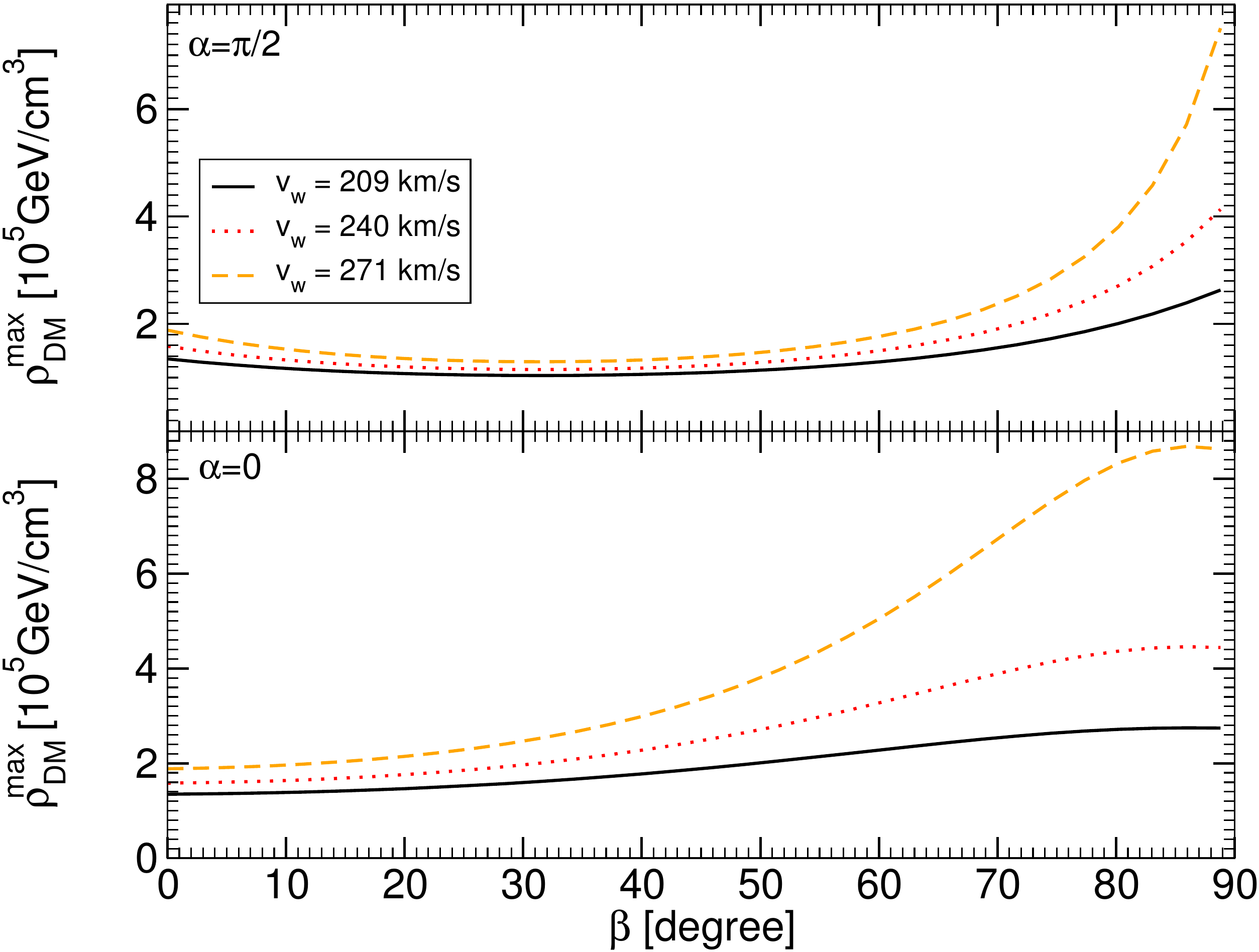,width=0.47\textwidth,angle=0,clip=true}
\caption{Upper bounds on the DM density due to DM dynamical friction as a function of the angle $\beta$ of the DM wind derived from the timing of binary pulsar J1713+0747. We adopted the observed values $P_b\approx 67.8\,{\rm day}$, $e\approx0$, $m_1\approx 1.31 M_\odot$, $m_2\approx 0.29 M_\odot$, $\iota\approx 75.3^\circ$, $\omega\approx176.2^\circ$ and considered $v_w\sim (240\pm 31)\,{\rm km/s}$. The upper (lower) panel shows the case with $\alpha=\pi/2$ ($\alpha=0$). The behavior is monotonic with $\alpha$ so these two cases bracket the minimum and the maximum upper bound.
\label{fig:BOUND}}
\end{center}
\end{figure}

By inserting these values into Eqs.~\eqref{PBDOTDF}, \eqref{PBDOTcm} and \eqref{PBDOTiota}, we can evaluate the total secular change of the orbital period and estimate the maximum value of $\rho_\DM$ such that $|\langle\dot P_b^\DF+\dot P_b^{\rm cm}+\dot P_b^\iota\rangle| \lesssim 2\times10^{-13}$, which is the value of $\dot P_b^\INT$ observed for J1713+0747~\cite{0004-637X-809-1-41}. This condition implies the upper bound on $\rho_\DM$ shown in Fig.~\ref{fig:BOUND}. 
From these results we obtain
\begin{equation}
 \rho_\DM\lesssim (1-8)\times 10^5\, {\rm GeV/cm}^3\,, \label{finalbound}
\end{equation}
where the boundaries of the range correspond to the maximum and minimum values of $\rho_\DM^\MAX$ in the range $\alpha\in[0,\pi/2]$, $\beta\in[0,\pi/2]$ and $v_w\in[209,271]\,{\rm km/s}$. As clear from Fig.~\ref{fig:BOUND}, the upper bound is only mildly dependent\footnote{We recall that $\dot P_b^\DF$ is independent of $\alpha$, so the mild dependence shown in Fig.~\ref{fig:BOUND} comes entirely from the two other terms, $\dot P_b^{\rm cm}$ and $\dot P_b^\iota$, and mainly from the latter.} on the angle $\alpha$ and displays only small variations in the entire range $\beta\in[0,\pi/2]$ and $v_w\in[209,271]\,{\rm km/s}$. Thus, the upper bound on $\rho_\DM$ is quite solid against the uncertainties on the three-dimensional velocity of this binary pulsar. 
Finally, the upper bound~\eqref{finalbound} scales linearly with $\dot P_b^{\rm xs}$ so any future improvement on the measurement of $\dot P_b^{\rm xs}$ for J1713+0747 would provide a more stringent constraint on $\rho_\DM$.

%

\section{Discussion}
The very fact that pulsar-timing techniques have reached the precision to be sensitive to the effect of DM in binary pulsars is certainly intriguing. Our results show that these effects are observable for various binaries in DM-rich environments.
DM dynamical friction introduces a peculiar annual modulation of the orbit, similar in spirit to the one expected for the scattering rate in DM direct-detection experiments on Earth~\cite{PhysRevD.33.3495,2012arXiv1209.3339F}. 

Current observations can be used to derive mild constraints on the local DM density, but are limited by the fact that none of the observed binary pulsars are located very close to the Galactic center, where the DM density is expected to peak. The next Square Kilometer Array (SKA) telescope~\cite{SKA} can improve the precision of current measurements by $2$ orders of magnitude, potentially probing DM densities lower than the DM density in the Solar System.  SKA is also expected to increase the number of observed pulsars by a factor of $10$, thus enhancing the chances of observing a binary pulsar in regions of high DM density.
Our analysis suggests that precise models of the effects of DM on the orbital motion would be mandatory to reach high accuracy in timing models for such binaries.

In this paper we presented an exploratory study that is valid only for binaries with long orbital period. For more relativistic binaries with $P_b\lesssim 0.6\,{\rm day}$, the standard computation of the gravitational-drag term~\eqref{DF} should be extended to take into account the effect of the wake of one object on the companion.

The constraints derived here apply to any form of (collisionless) DM and are therefore model independent. In fact, similar effects are also expected for the gravitational drag of ordinary matter in the interstellar medium. In that case, cohesion forces and matter compressibility should be taken into account and the corrections would significantly depend on how the relative velocity compares to the speed of sound in the medium~\cite{1999ApJ...513..252O,Macedo:2013qea,Barausse:2014tra}.

A more rigorous analysis requires us to include the effects discussed in this work in a timing model, an important task that we leave for future work. This might also be important to investigate the annual modulation of the orbital elements together with their secular changes. This peculiar annual modulation and the possibility of measuring $\ddot{P}_b$ in some systems can be used to discriminate the effect of DM from other mechanisms~\cite{HandbookPulsar,Doroshenko:2001zi,Lazaridis:2011wc,Pletsch:2015nia} that might cause comparable changes in the intrinsic orbital period.  A more sophisticated analysis can improve the constraints derived in this paper and would be crucial to devise detection tools, rather than to simply put upper bounds on the DM density. We hope that the prospects raised by our analysis will encourage further research in this direction.

\begin{acknowledgments}
We are indebted to Norbert Wex and Dick Manchester for very instructive correspondence, to Matteo Cadeddu, Vitor Cardoso and Leonardo Gualtieri for interesting discussions, and to two anonymous referees for several valuable comments.
This work was supported by the European Community through
the Intra-European Marie Curie Contract No.~AstroGRAphy-2013-623439, by FCT-Portugal through Project No.~IF/00293/2013, by the NRHEP 295189 FP7-PEOPLE-2011-IRSES Grant, and by the COST Action MP1304 ``NewCompStar''. 
\end{acknowledgments}

\appendix
\section{On the dynamical friction formula}\label{app:DF}
Chandrasekhar's dynamical friction formula for a single object moving through a homogeneous density distribution~\cite{Chandrasekhar:1943ys} is derived under a number of assumptions. Here we repeat the standard derivation and discuss the validity of certain assumptions.

Let us consider a star with mass $m_1$ and velocity $\mathbf{v}$ moving through a homogeneous density distribution made of light particles with mass $m_\DM\ll m_1$. Due to the encounter with a particle of mass $m_\DM$ moving with velocity $\mathbf{u}$, the velocity of the star is incremented by $\Delta v_\parallel$ and $\Delta \bm{v_\bot}$, in the directions parallel and perpendicular to $\mathbf{v}$, respectively. The total incremental velocity suffered by the star can be obtained by summing all contributions of the particles with mass $m_\DM$ over an interval of time which is much longer than the typical interaction time but also much shorter than the time scale over which $\mathbf{v}$ changes. As expected by symmetry arguments, when summed over a large number of encounters $\sum \Delta \bm{v_\bot}$ vanishes, while the incremental change in the direction of motion reads~\cite{RevModPhys.21.383}
\begin{equation}
 \frac{d\mathbf{v}}{dt}=-\pi \frac{(m_1+m_\DM)m_\DM G^2}{v^3}\mathbf{v}\int_0^\infty d^3u f(u) Q(u)\,, \label{DF0}
\end{equation}
where $m_\DM f(u)$ is the density distribution and
\begin{equation}
 Q(u)=\left\{ \begin{array}{ll}
               \log\left[(1+q^2(v+u)^4)(1+q^2(v-u)^4)\right] \,& u<v \\
               \log\left[(1+16q^2u^4\right]-4		     & u=v \\
               \log\frac{1+q^2(u+v)^4}{1+q^2(u-v)^4}-\frac{8v}{u} & u>v
              \end{array}\right.\,, \nn
\end{equation}
with $q=b_{\rm max}/[G(m_1+m_\DM)]$ and $b_{\rm max}$ being the characteristic size of the medium. Assuming $q^2(v\pm u)^4\gg1$, Eq.~\eqref{DF0} becomes~\cite{1991ApJ...379..280G}
\begin{eqnarray}
 \frac{d\mathbf{v}}{dt}&=&-4\pi \frac{ G^2 (m_1+m_\DM)m_\DM}{v^3}\mathbf{v}\nn\\
 &&\times\left[\int_0^v d^3u f(u)\left(\log(qv^2)+\log\frac{v^2-u^2}{v^2}\right)\right.\nn\\
 &&\left.+\int_v^\infty d^3u f(u)\left(-\frac{2v}{u}+\log\frac{u+v}{u-v}\right)  \right]\,. \label{DF1}
\end{eqnarray}
If the velocity dispersion $\sigma$ is large, $\sigma\gg v$, the above equation reduces to
\begin{eqnarray}
 \frac{d\mathbf{v}}{dt}&&\sim-4\pi \frac{G^2(m_1+m_\DM)m_\DM }{v^3}\mathbf{v}\nn\\
 &\times&\left[\log(qv^2)\int_0^v d^3u f(u)+\frac{2}{3}v^3	\int_v^\infty d^3u \frac{f(u)}{u^3}  \right]\,.\nn\\ \label{DF2}
\end{eqnarray}
Equation~\eqref{DF2} contains two contributions to the viscous drag on $m_1$, namely a term arising from slow particles with velocity smaller than $v$ and a term coming from fast particles with velocity larger than $v$. As noted in Ref.~\cite{1991ApJ...379..280G}, these two terms are roughly of the same order. For an isotropic, Maxwellian distribution $m_\DM f(u)=\frac{\rho_\DM}{(2\pi\sigma^2)^{3/2}}e^{-u^2/(2\sigma^2)}$, these two contributions combine to give
\begin{equation}
 \frac{d\mathbf{v}}{dt}\sim-\frac{4\sqrt{2\pi}}{3}\lambda \, \rho_\DM \frac{G^2(m_1+m_\DM)}{\sigma^3}    \mathbf{v}\,. \label{DF3}
\end{equation}
where $\lambda=\log(q\sigma^2)$ is the Coulomb logarithm. Equation~\eqref{DF3} coincides with the $\sigma\gg v$ limit of Chandrasekhar's formula~\cite{Chandrasekhar:1943ys,RevModPhys.21.383}
\begin{equation}
 \frac{d\mathbf{v}}{dt}=-4\pi \lambda\,\rho_{\DM}   \frac{G^2 (m_1+m_\DM) }{v^3}\left(\erf{(x)}-\frac{2x}{\sqrt{\pi}}e^{-x^2}\right){\mathbf{v}}\,,\label{DFv}
\end{equation}
where $x=v/(\sqrt{2}\sigma)$. 
Although it is usually considered that dynamical friction on a single body is due to ambient particles moving slower than the perturber, in fact Eqs.~\eqref{DF1} and ~\eqref{DF2} show that faster particles also contribute to the final result. 

In a binary system, the situation is actually opposite to the case of dynamical friction on a single perturber~\cite{1991ApJ...379..280G}. In that case the contribution from faster particles is dominant because such encounters have an impact parameter smaller than the orbital distance and interact with each component of the binary system separately. On the other hand, slower particles interact with the binary as a whole through tidal forces, only perturbing its center of mass~\cite{1991ApJ...379..280G,Quinlan:1996vp}.
Interestingly, for a Maxwellian distribution the two effects are equivalent but multiplied by two different Coulomb logarithms, one (arising from an integral over the impact parameter) is associated to slow encounters, whereas the other (arising from an integral over velocity space) is associated to fast encounters. In this case the final result is still described by Eq.~\eqref{DFv} with an effective Coulomb logarithm $\lambda$.
This property is a great advantage for our analysis and is valid only for Maxwellian or nearly Maxwellian distributions, as those assumed in the main text.
As noted in the main text our results, based on a superposition of Eq.~\eqref{DFv} for each single object of the binary system, are in exact agreement with those derived by Gould~\cite{1991ApJ...379..280G} in the large-$\sigma$ limit and within a completely different framework. This agreement gives further support to the validity of our analysis.

\begin{figure}[th]
\begin{center}
\epsfig{file=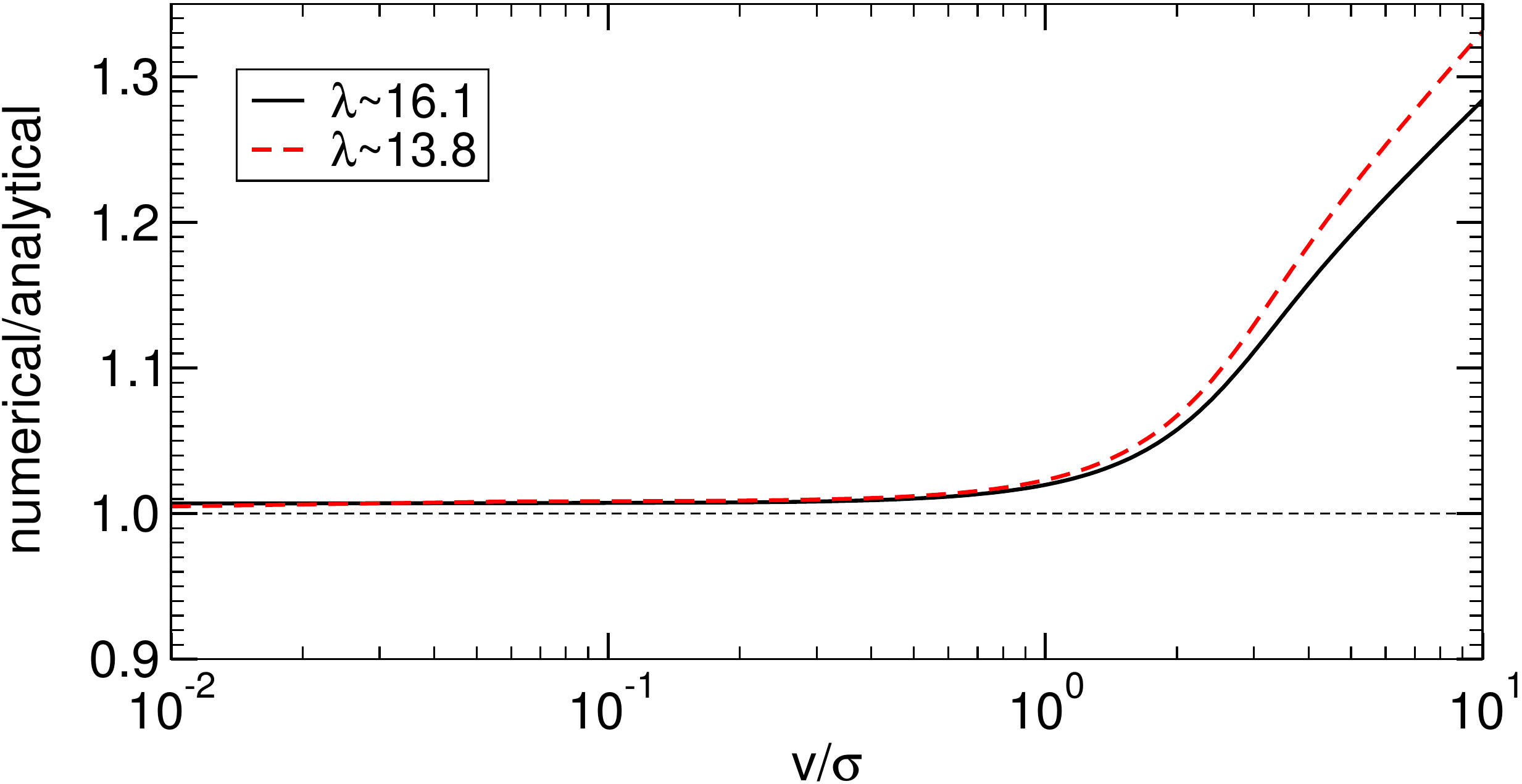,width=0.47\textwidth,angle=0,clip=true}
\caption{Ratio between Eq.~\eqref{DF0} and Eq.~\eqref{DFv} as a function of $v$. The agreement of the exact result with the analytical formula~\eqref{DFv} is very good for any $v<\sigma$.
\label{fig:comparison}}
\end{center}
\end{figure}

Nonetheless, the final formula~\eqref{DFv} is obtained from Eq.~\eqref{DF0} after various approximations~\cite{RevModPhys.21.383} whose validity is unclear (for example, while $q^2(v+ u)^4\gg1$ in typical situations, $q^2(v- u)^4$ is a small quantity when $u\sim v$). Thus, it is relevant to compare the exact numerical result against the approximate one typically used. We make this comparison in Fig.~\ref{fig:comparison}.
%
The analytical formula~\eqref{DFv} reproduces the exact result within a few percent or better for $v<\sigma$ and deviates at most by $\sim35\%$ when $v\sim 10\sigma$. In the main text we focus on the regime~\eqref{condPb}, where the analytical formula is in very good agreement with the exact results. Even for mildly relativistic pulsar binaries with $v\sim 2\sigma\sim 300\,{\rm km/s}$, Chandrasekhar's formula is accurate roughly within $6\%$.

\section{Computation of $\dot P_b^{\rm cm}$}\label{app:CM}
Our starting points are Eq.~\eqref{E2} and Eq.~\eqref{CM}. To first order, Eq.~\eqref{E2} can be written as $\dot{\mathbf{V}}=a_2\eta\mathbf{v}+a_3\mathbf{v}_w$. We decompose the DM wind vector as $\mathbf{v}_w=v_w\cos\alpha\cos\beta \mathbf{e}_x + v_w\sin\alpha\sin\beta \mathbf{e}_y + v_w \cos\beta\mathbf{e}_z$, where $\mathbf{e}_x$, $\mathbf{e}_y$ and $\mathbf{e}_z$ are the  unit vectors in the orbital frame.  The latter are related to the unit vectors $\mathbf{e}_X$, $\mathbf{e}_Y$, $\mathbf{e}_Z$ in the fundamental frame by a sequence of rotations (cf. Ref.~\cite{PoissonWill} for details). We further decompose $\mathbf{v}=v_n \mathbf{n}+ v_\lambda \bm{\lambda}+ v_z\mathbf{e}_z$. To zeroth order and in the circular case, $\mathbf{v}=v \bm{\lambda}$.

The final expression Eq.~\eqref{PBDOTcm} can be obtained by making use of the relations 
\begin{eqnarray}
 \bm{\lambda}\cdot \mathbf{e}_Z &=&\sin\iota \cos(\Omega_0t+\omega)\,,\\
 \mathbf{e}_x\cdot \mathbf{e}_Z &=&\sin\iota\sin\omega\,,\\
 \mathbf{e}_y\cdot \mathbf{e}_Z &=&\sin\iota\cos\omega\,,\\
 \mathbf{e}_z\cdot \mathbf{e}_Z &=&\cos\iota\,,
\end{eqnarray}
which are derived in Sec.~3.5.2 of Ref.~\cite{PoissonWill}.

\bibliography{biblio}
\end{document}